\documentclass[prl,aps,floatfix,twocolumn,showpacs]{revtex4}
\usepackage{latexsym,amssymb,graphicx,caption}
\usepackage{epsfig}

\newcommand{\be}{\begin{equation}}
\newcommand{\ee}{\end{equation}}
\newcommand{\bea}{\begin{eqnarray}}
\newcommand{\eea}{\end{eqnarray}}

\def\figone#1#2#3{\begin{figure}
\centering \leavevmode
\epsfxsize=0.8\columnwidth \epsfbox{#1}
\caption{#2 \label{#3}}
\end{figure} }

\begin{document}

\title{Jamming as a critical phenomenon: a field theory of zero-temperature grain packings}

\author{Silke Henkes and Bulbul Chakraborty }

\affiliation{Martin Fisher School of Physics, Brandeis University,
Mailstop 057, Waltham, Massachusetts 02454-9110, USA}

\begin{abstract}
A field theory of frictionless  grain packings in two dimensions is shown to exhibit a zero-temperature critical point at a non-zero value of the packing fraction.  
The zero-temperature constraint of force-balance plays a crucial role in determining the nature of the transition.   Two order parameters, $<\!z\!>$, the deviation of 
the average number of contacts from the isostatic value and, $<\!\phi\!>$,  the average magnitude of the force per contact, characterize the transition from the jammed 
(high packing fraction) to the unjammed (low packing fraction state).
The critical point has a mixed character with the order parameters showing a jump discontinuity but with fluctuations of the contact force diverging.   
At the critical point, the distribution of $\phi$ shows the characteristic plateau observed in static granular piles.
 The theory makes  falsifiable  predictions about the spatial fluctuations of the contact forces.
Implications for finite temperature dynamics and generalizations to frictional packings and higher dimensions are discussed.
\end{abstract}


\maketitle

\paragraph{Introduction}

In a remarkably diverse range of systems,  the transition from a flowing, liquid state  to  a jammed, solid state is heralded by a dramatic slowing
down of relaxations\cite{jammingreview}.
Does an equilibrium critical point underlie this glassy dynamics ?  
The debate surrounding this question has been spurred by the absence of any obvious static signature accompanying the rapid increase 
of time scales\cite{jammingreview}. 
Purely dynamical scenarios 
have been proposed\cite{biroli} to explain time-scale divergences with no accompanying static divergences.   For thermal systems,  a different perspective 
has been offered within the framework of an avoided critical point\cite{kivelson} 
and a scaling theory based on the existence of a zero-temperature critical point\cite{sethna}.   In a more recent development, it 
has been suggested that the mechanism of jamming in both thermal and athermal systems is controlled by a zero-temperature critical point (J-point)\cite{jamminglong}. 

Experiments on weakly sheared granular media indicate that at a critical packing fraction there is a transition accompanied by slow dynamics, vanishing of mean stress, increasing stress
fluctuations and a change in the distribution of contact forces\cite{stressfluc1,ericflow}.  Simulations  indicate a critical point 
occurring at zero temperature and a packing fraction close to the random  close packing  value\cite{jamminglong}.
At this critical point the grain packing is isostatic, having reached the special coordination where all contact forces are completely determined by the packing geometry\cite{jammingreview}.  
A theory based on this observation predicts a diverging length scale 
associated with the mechanical stability of the network\cite{wyart}.  In a different 
theoretical approach, an analogy has been drawn between the jamming transition and  k-core percolation\cite{schwartz}.  

In the present work, it is shown that a statistical field theory of two-dimensional,  zero-temperature, frictionless grain packings exhibits  a critical point. This point separates a disordered phase from an ``ordered'' one 
characterized by two order parameters: (i) the magnitude of the force per contact, $<\!\phi\!>$, and (ii)  $<\!z\!>$, the deviation of the contact number per grain from  its isostatic value.
At the critical point, the fluctuations around $<\!\phi\!>$ diverge but those around $<\!z\!>$ go to zero.
An analytic prediction for  $P(F)$,  the distribution of contact forces,  is in excellent agreement with experiments\cite{stressfluc1}, and the theory makes falsifiable  predictions regarding the spatial correlations of the forces. 

\paragraph{Statistical Ensemble} In granular matter, which is athermal, a natural control parameter is the packing fraction\cite{jammingreview}. 
For short-range repulsive potentials,  the corresponding statistical ensemble
is one with a fixed average pressure\cite{foot1}.
The probability  $P[\{ {\bf r}_i \}]$ of  a grain packing with the set of positions $\{ {\bf r}_i \}$ can be 
obtained by using the maximum entropy principle\cite{jaynes}, which also forms the basis of the Edwards ensemble of granular packings\cite{edwardsmounfield}.  
Maximizing the entropy, $S[P]=-\sum_{\{ {\bf r}_i \}}P[\{ {\bf r}_i \}]\ln{P[\{ {\bf r}_i \}]} $ subject to the constraint of fixed average pressure leads to\cite{chandlerbook}:
\begin{equation}
P[\{ {\bf r}_i \}] = (1/Z) \exp (\alpha p(\{ {\bf r}_i \}))
\label{prob}\end{equation} where $p(\{ {\bf r}_i \}) =(1/V) {\sum}_{ij}
 r_{ij}  \,{\partial U \over \partial r}|_{r_{ij}}$ is the pressure
of the configuration  $\{ {\bf r}_i \}$( $U$ is the interaction potential). The ``canonical'' partition function, 
$Z(\alpha)={\sum}_{{\bf r}_i}^{\prime} P[\{ {\bf r}_i \}] $  is the generating function of all statistical averages.
The Lagrange multiplier $\alpha$ plays the role of inverse temperature: $\alpha = -\partial S/\partial\!\!<\!p\!>$ and the prime
on the summation restricts it to grain configurations satisfying the equations of mechanical equilibrium for frictionless packings in $d$ dimensions:
\begin{equation}
{dN~\rm eqs}:~~ \sum_{j} F_{ij} {{\bf r}_{ij} \over |{\bf
r}_{ij}|} =0 \label{forceb}\end{equation} 
\begin{equation}
{ <\!\!z\!\!>N/2 ~\rm eqs}:~~ F_{ij} = f({\bf r}_{ij})
\label{forcelaw}\end{equation} 
Here $<\!z\!>$ is the average number of  contacts per grain, $F_{ij}$ is the magnitude of the
contact force between  grains $i$ and $j$, and $f({\bf r}_{ij})$ is the function specifying
the  inter-grain force law.    At the isostatic point,  $<\!z\!>=z_{iso}=2d$, 
the number of equations in Eq. \ref{forceb} is exactly equal to the number of unknown forces, $F_{ij}$\cite{wittentkachenko,vanhecke} and, therefore,  the forces are uniquely determined 
by these equations.  The positions $\lbrace {\bf r_{i}} \rbrace $ can then be obtained by inverting Eq. \ref{forcelaw}.
For $<\!\!z\!\!>\,  > \, z_{iso}$,  Eqs. \ref{forceb} and \ref{forcelaw} are coupled and the coupling can be parametrized by $\epsilon$ which encodes how
sensitive the contact forces are to changes in the grain
positions \cite{vanhecke}:
\begin{equation}
\epsilon = {<\!F\!> \over <\!r_{ij}\!>} <{dF_{ij} \over
dr_{ij}}>^{-1}\,.\label{epsilon}\end{equation}  
For hard spheres, $\epsilon=0$, 
the two equations are
decoupled and the only packings for which $\{F_{ij}\}$'s and $\{ {\bf r}_{ij}\}$'s can be determined
are the isostatic ones\cite{wittentkachenko}.  For  $\epsilon << 1$, 
small variations in grain positions can lead to force changes
comparable to the average force and Eqs. \ref{forceb} and \ref{forcelaw} are weakly coupled.  
An effective way of calculating the partition function, in this limit,  is to  sum over all the  solutions, $\{F_{ij}\}$,
to Eq. \ref{forceb} for a given set $\{ {\bf r}_i \}$ and impose a Gaussian constraint; $e^{-(\epsilon/2){\sum}_i(p_i -
p^0_i)^2}$.  Here $p_{i}$ is the
pressure calculated from  $\{{\bf r_{i}}\}$ and $p_{i}^{0}$ is calculated from $\{F_{ij}\}$:
\begin{eqnarray}
p_i &=& {\sum}_j r_{ij} f(r_{ij}) \nonumber \\
p_{i}^{0} &=&{\sum}_j r_{ij} F{ij} \nonumber \\
Z(\alpha)& =&{\sum}_{\{{\bf r}_{i}\} ,\{ {F_{ij}}\}}  e^{(\alpha {\sum}_i
p_i)}e^{-\epsilon /2 {\sum}_i (p_i -p^0_i)^2} \,.\label{gaussian}\end{eqnarray} 
\figone{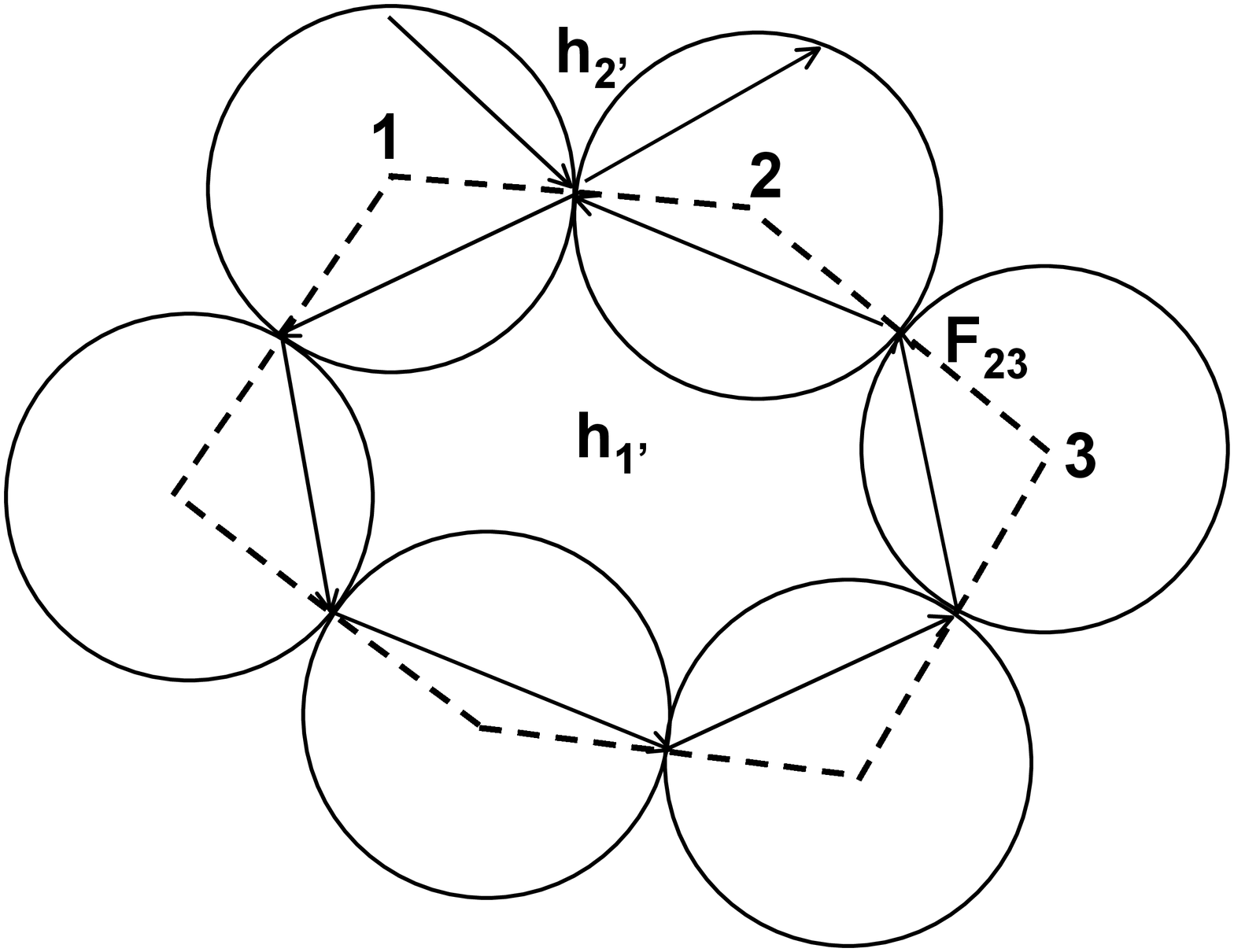}{Illustration of height field.  The dashed lines are contact forces
and the heights associated with the loops are represented by the arrows encircling a loop }{loopforces}
The sum over $F_{ij}$ is {\it constrained} to only
those forces which satisfy force-balance.    These constraints can be  incorporated  through the introduction of  a set of auxiliary variables, the loop forces\cite{ballblumenfeld}. 
Loop forces  or `heights', $\{{\bf h}_{i}\}$,\cite{ballblumenfeld} are vectors associated with the the voids enclosed by grains.
The height vectors are related to the the contact forces through: ${\bf F}_{ij}= {\bf h}_{j'} - {\bf h}_{i'}$ where $j'$ and $i'$ are the voids bracketing
the contact $ij$ ({\it cf}  Fig. \ref{loopforces}).  Since the ${\bf F}_{ij}$'s  around a grain sum to zero,  the mapping of forces to heights is one to one, 
up to an arbitrary choice of a single height.    For frictionless grains,  the requirement of ${\bf F}_{ij}$ being  parallel to ${\bf r}_{ij}$  leads to restrictions on the heights.
In developing the field theory, the heights are coarse grained over a mesoscopic region with a length scale much larger than the grain radius but smaller than 
a typical length scale over which the height fields vary\cite{ballblumenfeld} to define a continuous height field.
It can be shown\cite{shenkes} that the constraint on the coarse-grained field resulting from the frictionless property is that the height field be divergence free and, in  $2d$, this implies that
the height field can be written in terms of a scalar potential\cite{ballblumenfeld} $\psi$:
$
h_{x} = \partial_{y}\psi ~; h_{y}=-\partial_{x}\psi$.  A similar approach led to the definition of a potential for frictional grain-packings\cite{ballblumenfeld}.   
The stress tensor, coarse grained over a mesoscopic region of area $A$,  around the point $\bf r$: $\hat \sigma^{0}({\bf r})= (1/A) {\sum}_{j  \subset A}{\sum}_k {\bf r}_{jk} {\bf F}_{jk}$, 
is expressed in terms of $\psi$ as\cite{ballblumenfeld}
$$
\hat \sigma^{0}  = \left [\begin{array}{cc}
\partial_{y}^{2}\psi & -\partial_{x}\partial_{y}\psi \\
-\partial_{x}\partial_{y}\psi& \partial_{x}^{2} \psi\\
\end{array} \right]\,,$$
and the coarse grained pressure is, $p^{0}({\bf r})  \equiv Tr \hat \sigma = {\nabla}^2 \psi ({\bf r}) $.   The partition function can, therefore, be written in terms of the {\it unconstrained} field $\psi$
 
The coarse-grained field $\psi$  carries with it a weight $\Omega[\psi]$ which counts
the number of microscopic $\lbrace F_{ij} \rbrace$
configurations giving   rise to the same $\psi$
field.  Arguments similar to the ones employed in height-maps
of loop models\cite{janedimer},  lead to $ \Omega[\psi] \sim e^{-(1/2) \int d^{d} r ({\nabla}^2 {\psi})^{2}}$.  The
basic reasoning is that configurations which are completely flat,
i.e. ${\nabla}^2 \psi = 0$, have the largest number of loops (dashed in Fig. \ref{loopforces}) along
which the contact forces can be shuffled without violating the
force-balance constraint and, therefore, they are favored entropically.

At the isostatic point, $z=z_{iso}$,  $p = {\nabla}^2 \psi$.  For packings close
to the isostatic one,  an expansion in $z-z_{iso}$ leads to 
\begin{eqnarray}
p({\bf r})& =& {\nabla}^2 \psi+ (z_-z_{iso}) {\Delta p \over \Delta z}|_{z_{iso}}\nonumber \\
&=&{\nabla}^2 \psi + (z-z_{iso}) {{\nabla}^2 \psi \over z_{iso}} = z {{\nabla}^2 \psi \over z_{iso}} ~.\label{pressure}\end{eqnarray}
The second set of equations follow by noting that the extra pressure, ${\Delta p \over \Delta z}|_{z_{iso}}$,  due to the introduction of an additional contact at the isostatic point
is the pressure at this packing divided by the number of contacts.   Just as the mapping from $F_{ij} \rightarrow \psi$,  
the mapping from 
${\bf r_{i}}$ to the coarse-grained pressure, $p({\bf r})$, involves the calculation of a weight  $\omega[p]$.   
It can be argued  that  $\omega[p] \sim \exp (-K |({\bf \nabla} p)|^2)$ 
by picturing the packing of $N$ grains in a volume $V$.
Entropically, it is favorable to have the grains occupy as much of the volume as possible, leading to 
small spatial variations of the packing density and $p(\bf r)$.  Large pressure gradients occur in the entropically unfavorable  packings 
where there are large variations in the packing density.

With these weights,   the partition function is:
\begin{eqnarray}
Z&=& {{\sum}_{\{z\}, \{\psi\}}}\Omega[\psi]\omega[p] 
\{e^{{\alpha}\!<\!z\!>\int d^{d}r p(r)}\}\nonumber \\
&&
\{e^{-\epsilon /2 \int d^{d}r (p(r) -p^{0}(r)))^2}\} \nonumber \\
&&={\sum}_{\{z\}, \{\psi\}}\ e^{-H[{\psi},z]} \nonumber \\
&&H =  {\sum}_{\bf q} ({1 \over 2} |\phi_{{\bf q}}|^2
-\alpha (z_{iso}\delta_{\bf q} + z_{-{\bf q}} ){\phi}_{\bf q})\nonumber\\
&+ & V^{-1}{\sum}_{{\bf q}_1 ,{\bf q}_2, {\bf q}_3 ,{\bf q}_4}
[{\epsilon \over 2} - {K \over 2} ({\bf q}_1 + {\bf q}_3)\cdot
({\bf q}_2 + {\bf q}_4)] \nonumber\\
&&[z_{{\bf q}_1}z_{{\bf q}_2}{\phi}_{{\bf
q}_3}{\phi}_{{\bf q}_4} \delta_{{\bf q}_1+{\bf q}_2+{\bf q}_3+{\bf
q}_4}]\label{partition}\end{eqnarray}
Here,  ${\phi}= {\nabla}^2 {\psi}$ represents the magnitude of the force per contact ({\it cf} Eq. \ref{pressure}), and the field $z$  has been redefined to $z-z_{iso}$ which is restricted to the set of integers.
The parametes $\alpha$, $\epsilon$  and $K$
have been scaled to absorb resulting factors of $z_{iso}$.  

\paragraph{Critical Point}The Hamiltonian $H$ reflects the competition between $\Omega[\psi]$ favoring $\phi =0$ and the "field" $\alpha$ favoring non zero $\phi$.
It provides a model for studying the response functions of grain packings with small but  finite $\epsilon$.
We assume that the  integer restriction on $z$ can be  ignored\cite{chuiweeks} as long as $\epsilon \ne 0$

In investigating whether or not  there is a finite-$\alpha$ phase transition involving the vanishing of one or more order parameters, the fields in the Hamiltonian in Eq.
\ref{partition} are expanded around their average values; ${\phi}_{\bf q}
= <\!\phi_{\bf q}> + {\zeta}_{\bf q}$, $z_{\bf q} = <z_{\bf q}> +
{\eta}_{\bf q}$ and the hamiltonian written as $H = H_{0}(<\!\phi_{q}\!>,<z_{q}\!>) + H_{1}(<\!\phi_{q}\!>,<z_{q}\!>;\zeta_{q},\eta_{q})$. The  order parameters $<\!\phi_{q}\!>$ and $<z_{q}\!>$  are obtained 
 by minimizing the effective  potential, $\Gamma (<\!\phi_{q}\!>,<z_{q}\!>) = H_0
(<\!\phi_{q}\!>,<z_{q}\!>) - \ln (\int {\Pi}_{\bf q} d{\zeta}_{\bf q} {\Pi}_{\bf
q} d{\eta}_{\bf q} e^{-H_1})$.
The
simplest, non-trivial approximation, is obtained by
calculating  the fluctuations, 
$<|{\zeta}_{\bf q}|^2>$ and $<|{\eta}_{\bf q}|^2>$,  at the loop level,  replacing all 4-point averages by 2-point averages\cite{amitbook},  and assuming spatially
uniform order parameters: $<\!\phi\!>\equiv<\!\phi_{q=0}\!>$,
$<\!z\!>\equiv<\!z_{q=0}\!>$. To  leading order in $\epsilon$:
\begin{eqnarray}
<|{\zeta}_{\bf q}|^2>^{-1}&=& 1 +\epsilon <\!z\!>^{2}-1/<\!\phi\!>^{2}+K<\!z\!>^{2}q^{2}\nonumber \\
<|{\eta}_{\bf q}|^2>^{-1}&=& \epsilon <\!\phi\!>^{2}+K<\!\phi\!>^{2}q^{2}
\label{contactstruc}\end{eqnarray}
Incorporating these results in the effective potential and minimizing with respect to the order parameters leads to: 
\begin{eqnarray}
&&<\!\phi\!>-\alpha (z_{iso}+<\!z\!>) +\epsilon <\!z\!>^{2}<\!\phi\!> + {1 \over <\!\phi\!>} =0\nonumber \\
&&\epsilon (<\!\phi\!>^{2} + {1 \over {1-1/<\!\phi\!>^{2}}})<\!z\!> -\alpha <\!\phi\!> = 0 \label{meanfield}\end{eqnarray}
Solving these equations  near  $\alpha_{c } = 2/z_{iso}$,  the order parameters behave as:
\begin{eqnarray}
<\!\phi\!> &= &(\alpha /\alpha_{c}) (1+(1-\alpha_{c}/\alpha)^{1/2})\nonumber \\
<\!z\!> &=& (\alpha/\epsilon) (1-\alpha_{c}/\alpha)^{1/2}
\label{mfsolutions}\end{eqnarray}
For $\alpha \ge \alpha_{c}$, there is an ordered phase characterized by two  order parameters.   For $\alpha < \alpha_{c}$, 
 $\Gamma(<\!\phi\!>,<\!z\!>)$ ceases to have any local minima or maxima ({\it cf} Fig. \ref{P(F)}) and $<\!\phi\!>$ jumps discontinuosly to $0$: the physical limit of its allowed values.  
From  Eqs. \ref{contactstruc} and \ref{mfsolutions},  as $\alpha\! \rightarrow \!\alpha_{c}$,  $\phi \!\rightarrow \!1$  and the $q=0$ force fluctuations diverge: $<\!\phi^{2}\!> -<\!\phi\!>^{2} \sim (1-\alpha_{c}/\alpha)^{-1/2}$.  
This type of transition is indicative of the  end of a line of metastable 
 equilibrium similar to spinodal critical points.  Unlike spinodals, however,  the transition is accompanied by the disappearance of {\it any} local minimum;  
 a phenomenon observed  in models with rigid constraints such as certain dimer models\cite{hui}. The fluctuations of $z$ vanish as $<\!z^{2}\!>\!-\!<\!z\!>^{2} \sim (1-\alpha_{c}/\alpha)^{1/2}$.


The structure factor of the magnitude of the contact forces (the $\phi$ field),  $S({\bf q}|) \equiv  <|\zeta_{\bf q}|^2> $, provides detailed, measurable information
about the spatial fluctuations of the contact force.   Eq.   \ref{contactstruc} predicts
$$
S(q)  = {1 \over {{\xi}/a(\epsilon,K) +  q^2 {\xi}^{2}}}
$$
where $\xi \simeq  {(1-1/<\!\phi\!>^{2})}\simeq
(1-{\alpha_c}^2/{\alpha}^2)^{1/2}$ and $a(\epsilon,K)$ is a characteristic length scale of the model.    The length scale $\xi$  vanishes  at the critical point and  $S(q=0)$  diverges as $1/\xi$.
 There is another length scale related to the width of the structure factor peak:  
 $l^{-1} \equiv q_{0}$ where $q_{0}= 1/\sqrt{a\xi}$.  The width diverges as $(1-{\alpha_c}^{2}/{\alpha}^2)^{-1/4}$.   The appearance of these two length scales, and
 the specific form of $S(q)$, are predictions of the theory that should be testable in experiments and simulations.  The length scales, $l$ and $\xi$,
 are associated with correlations of contact forces and their  vanishing  implies  that the length scale over which the grain packing behaves as an elastic solid is going to zero or, equivalently,
 the length scale over which the packing is floppy is diverging.
 The two exponents, $1/2$ and $1/4$, associated with the divergence of these complementary length scales are
 reminiscent of the exponents discussed in theories and simulations of J-point\cite{jamminglong,wyart,schwartz}.  
 \figone{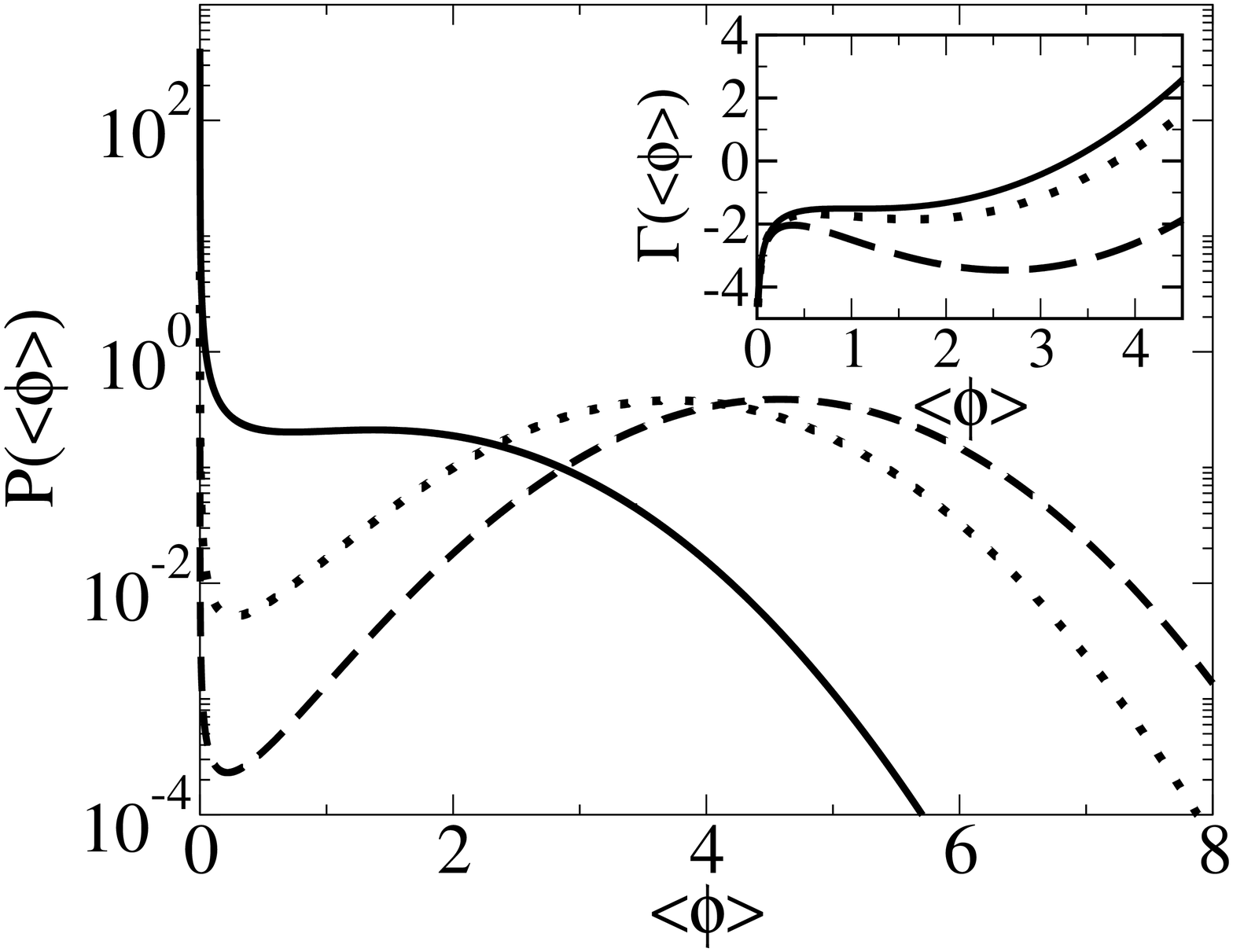}{The distribution  $P(<\!\phi\!>)$ of contact forces (from bottom to top) $\alpha -\alpha_{c}=0.7$, 
 $0.5$ and $0.025$.   The
 inset shows the $<\!z\!>=0$ cut of the  potential, $\Gamma(<\!\phi\!>,<\!z\!>)$ at different $\alpha$ with $\alpha \rightarrow \alpha_{c}$ 
 from bottom to top. }{P(F)}
 \paragraph{Distribution of contact forces}  Experiments\cite{stressfluc1,ericflow} and simulations\cite{jamminglong,vanhecke} 
 have  shown that changes in the distribution of $P(F)$ are associated with transitions involving the vanishing of stress.
 In the field theory, $<\!\phi\!>$  corresponds to $F$ and the distribution,  
 $P(<\!\phi\!>) \simeq e^{-\Gamma(<\!\phi\!>,<z(<\!\phi\!>))}$.
  The critical regime has $<\!z(<\!\phi\!>))\!> \simeq 0$   ({\it cf} Eq. \ref{meanfield}) implying 
  $\Gamma(<\!\phi\!>) \simeq (1/2) <\!\phi\!>^{2} - \alpha z_{iso} <\!\phi\!> + \ln(<\!\phi\!>) $.  The $\ln(<\!\phi\!>)$ term results from integrating out the $z$ field and
  embodies the physical effect of large contact number fluctuations for small  contact forces.  
  This term leads to a power-law decay $P(<\!\phi\!>) \sim 1/<\phi>$ near  $\phi = 0$ and provides a qualitative explanation of the experimental data\cite{stressfluc1}.
  Fig. \ref{P(F)} clearly demonstrates that the approach to $\alpha_{c}$
  is accompanied by changes in the low-force regime of $P(F)$ with a peak giving away to a plateau as the force fluctuations diverge.  Since  diverging force
  fluctuations imply vanishing of the shear modulus $\sim 1/(<\!\phi^{2}>-<\!\phi\!>^{2})$ , the shape change is connected to the disappearance of yield stress.  
  The theory does not yield the exponential tail of the force distribution which is not unexpected in a coarse-grained model since the large force
  behavior results from a stress redistribution at the particle level\cite{stressfluc1,qmodel}.  
  
  \paragraph{Conclusions}A field theory, founded on general properties of $T=0$ packings of frictionless grains in two-dimensions,  exhibits a
  critical point separating  a jammed, ``ordered''  phase with finite yield stress at large packing fractions ($\alpha >\alpha_{c}$) from an unjammed phase where
  the order parameters vanish.  A hallmark of the transition is a change in the distribution of contact forces reflecting diverging force fluctuations.   Two length scales, associated with 
  the spatial fluctuations of the contact forces, are predicted to go to zero at the critical point and should be detectable in experimental measurements of  the structure factor of  contact forces. 
 
  If the zero-temperature critical point controls the 
  low temperature and weakly-driven behavior, then glassy dynamics follows from general scaling arguments\cite{sethna,satya}.
  At finite temperatures, the force-balance constraint is violated and the height-field has defects\cite{janedimer}.  The interactions between these defects determine the stability of the zero-temperature
  critical point and, therefore, it will be crucial to understand these interactions.   
  The extension of the theory to frictional packings is within reach  since the loop-force formalism exists\cite{ballblumenfeld}.   
  Although the  similarities between two and three dimensions,  observed in  simulations,  suggest that an extension  to higher dimensions is possible,  this remains an open question.
  
  The authors would like to thank Jane' Kondev for many insightful comments including the suggestion about the entropy associated with a coarse-grained pressure field.  B.C. would
   like to acknowledge useful discussions with Tom Witten, Sid Nagel, Andrea Liu,  Leo Silbert and Matthew Wyart and 
   the hospitality of the James Franck Institute.  This work was supported by  the  NSF grants DMR-0207106 and DMR-0403997.
 

\begin{thebibliography}{0}
\bibitem{foot1}{The simulations of J-point (Ref 5) were peformed in an
ensemble where the pressure is controlled,  a "microcanonical" ensemble.}
\end{thebibliography}
 
\end{document}